\documentclass[epj]{svjour}
\usepackage{amsmath}
\usepackage[dvips]{graphicx}

\begin{document}

\title{Pulse pair generation from coherently prepared atomic ensembles}

\author{D. Moretti, D. Felinto, and J.~W.~R. Tabosa}

\institute{Departamento de F\'{\i}sica, Universidade Federal de
Pernambuco, 50670-901 Recife, PE - Brazil}

\date{\today}

\abstract{
We report a detailed investigation on the generation of pulse pairs during the readout of a coherence grating stored in a cold atomic ensemble. The pulse shapes and the split of the retrieved energy between the two pulses are studied as a function of the relative intensities of the two reading fields, and a minimum is observed for the total retrieved energy. We introduce a simplified analytical theory for the process, considering a three-level atomic system, which explains all the most striking experimental features. 
}

\maketitle

\section{Introduction}

In recent years the techniques for generation, storage, and
manipulation of light pulses by atomic systems have developed
rapidly in connection to the interest in constructing more general
quantum networks for various applications in quantum and classical
information and metrology \cite{Kimble08}. The idea is to combine the photon's capability for fast exchange of information with the long lived atomic memories in order to allow information to be processed in a distributed manner among various sites. In this field, the studies of light storage based on the phenomenon of Electromagnetic Induced Transparency (EIT) plays a central role~\cite{Harris97,Lukin03,Fleischhauer05}. Soon after the
theoretical proposal by Fleischhauer et al \cite{Fleischhauer00} for such
light storage in an atomic ensemble, the demonstration of this
phenomenon was reported by Liu et al \cite{Liu01} and Phillips et al
\cite{Phillips01}. Since then these demonstrations have triggered a
number of related works reporting different aspects of the light storage
process \cite{Zibrov02,Wang05,Yu05,Tabosa07,Moretti2008}.

Parallel to the development of the storage of light pulses by atomic ensembles came various proposals for using such effect in quantum information applications. The controllable storage of single photons, for example, could be a great step in the implementation of linear optics quantum computation~\cite{Knill2001}. The use of atomic ensembles to generate and store single photons could also result in large efficiency gains for the distribution of entanglement over long distances, as in the DLCZ protocol for quantum repeaters~\cite{Duan2001}. Presently, prove-of-principle experiments have already demonstrated both single photon storage~\cite{Eisaman2005,Chaneliere2005} and various building blocks of the DLCZ protocol~\cite{Chou2005,Chou2007}. Even though such experiments still present efficiencies and coherence times considerably lower than the ones required for the actual implementation of the above mentioned quantum algorithms, they demonstrate the versatility and power, for various applications, of the process of light storage by atomic ensembles.

Our present work explores new possibilities of such light storage process involving the extraction of the stored information in different directions and the possibility to distribute the information in more than one optical mode at once. In our scheme, an ensemble of cold atoms is initially prepared in a coherent superposition of Zeeman ground states by a pair of writing beams with a small angle between their respective propagation directions. After the turn off of these beams a coherence grating is left in the ensemble, storing information related to various aspects of the optical state of the writing fields~\cite{Moretti2008,Moretti2009}. In the present work, we only use laser pulses as writing fields. However, in order to understand the implications of our work, it is important to have in mind that this configuration for writing optical information in an atomic ensemble is the same as the ones employed for both single-photon~\cite{Chaneliere2005} and squeezed-vacuum~\cite{Honda2008} storage in cold atoms. 

We emphasize here that the readout of the stored information can be manipulated in a quite general way, by both simple control of the direction of the retrieved pulse or the distribution of the stored information among different optical modes. In order to investigate such processes we extract the stored coherence grating using an independent pair of counter-propagating reading beams. In this way, by controlling the relative intensity of the two read fields we can choose between the extraction of the coherence grating as a single optical pulse in one particular direction, as a single optical pulse in the opposite direction, or as a pair of counter-propagating light pulses of arbitrary relative amplitudes. Such pulse pair generation from a stored coherence grating was first reported in Ref.~\cite{Tabosa07}. Here we approach this process under a more general framework, investigating the transition between the more well known processes of Delayed-Four-Wave-Mixing~\cite{Moretti2008,Matsukevich2005} and EIT~\cite{Liu01,Phillips01} readouts and such pulse pair generation. We also introduce a simple, analytical theory that accounts for all major features of this double-field readout, and use it to discuss various aspects of the process.

In the following, Section~\ref{theory} introduces the general idea of our scheme and the theory to model it, together with the analysis for various aspects of the pulse-pair generation process. Section~\ref{experiment} describes our experiment and its main results. It also provides the comparison between experimental results and the theory of Sec.~\ref{theory}. Such comparison reveals a quite satisfactory agreement, particularly if we keep in mind the relative simplicity of the theoretical model. Finally, in Sec.~\ref{conclusions} we provide a summary of our results and the perspective for future works.

\section{Theory}
\label{theory}

As mentioned above, we consider here an ensemble of cold atoms sequentially excited by two
pairs of laser fields, see Fig.~1. The first pair is composed by the
two writing fields $W$ and $W^{\prime}$, and the second pair by the
two reading fields $R$ and $R^{\prime}$. The atoms have a lambda
configuration for the relevant level structure, with two degenerate
ground states,$|1a\rangle$ and $|1b\rangle$, and one excited state
$|2\rangle$. These states correspond to specific Zeeman sublevels of
two hyperfine states, and thus the optical fields connecting them
have specific circular polarizations. The polarization of fields $W$
and $R^{\prime}$, connecting states $|1a\rangle$ and $|2\rangle$, is
then $\sigma^+$. Fields $W^{\prime}$ and $R$, on the other hand,
have $\sigma^-$ polarization and connect levels $|1b\rangle$ and
$|2\rangle$. These polarizations are defined in the reference frame
of the atom. Fields $W$ and $W^{\prime}$ have an angle $\theta$
between their respective propagation directions. The angle is small
enough for us to neglect any relative polarization change of the
beams due to this angle. Fields $R$ and $R^{\prime}$ have opposite
directions, but propagate in the same line as $W$.

The pair of writing fields act on the atomic ensemble for a long
time before they are turned off, leaving a coherence grating printed
in the ground state of the ensemble (Fig. 1a). This first
stage of coherent preparation of the system is discussed in
Sec.~\ref{preparation}. After some storage time $t_s$, the two read
fields are turned on, resulting in the emission of two other fields,
$D$ and $D^{\prime}$ (Fig. 1b). These two new fields are
generated as pulse pairs with relative intensities given by the
relative powers of the two reading fields. They carry the
information about the coherence grating originally stored in the
atomic ensemble. This reading process will be discussed in detail in Sec.~\ref{generation}.

\subsection{Coherent Preparation}
\label{preparation}

The coherent preparation process that we consider here was
introduced and discussed in detail in Ref.~\cite{Moretti2008}.
In this section, then, we will simply review these previous results
and prepare the necessary notation for the following sections. In
this way, we consider fields $W$ and $W^{\prime}$ as plane waves
propagating in the directions specified by the wave-vectors
$\vec{k}_W = k_W \hat{z}$ and $\vec{k}_{W^{\prime}}$, respectively,
with electric fields given by
\begin{subequations}
\begin{align}
\vec{E}_W &= {\cal E}_W e^{i(k_W z - \omega_W t)} \hat{\sigma}^{+} \;, \\
\vec{E}_{W^{\prime}} &= {\cal E}_{W^{\prime}}
e^{i(\vec{k}_{W^{\prime}}\cdot\vec{r} - \omega_{W^{\prime}} t)}
\hat{\sigma}^{-} \;,
\end{align}
\label{campo}
\end{subequations}
where ${\cal E}_W$ and ${\cal E}_{W^{\prime}}$ represent the
amplitude of each field, and $\omega_W$ and $\omega_{W^{\prime}}$
their frequencies. As discussed above, vectors $\vec{k}_W$ and
$\vec{k}_{W^{\prime}}$ form a small angle $\theta$.

\vspace{2.2cm}
\begin{figure}[ht]
\hspace{0.0cm}\includegraphics[width =8.5 cm,angle=0]{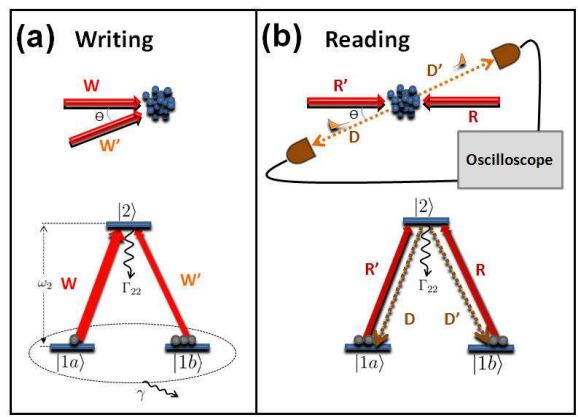}
\vspace{-2.7cm}
\caption{(colors online) Schematic representation of the (a) writing and (b) reading processes. Top: propagation directions of the writing ($W$,$W^{\prime}$), reading ($R$,$R^{\prime}$), and diffracted ($D$,$D^{\prime}$) fields. Base: level structure considered in the theory, with indication of the energy levels connected by each field. $\Gamma_{22}$ and $\gamma$ are the relaxation rates of the excited state and of the ground-state coherence, respectively.} 
\label{fig1}
\end{figure}

We are interested in the situation where the rate $\gamma$ of
decoherence between the ground states is much smaller then the
spontaneous decay rate $\Gamma_{22}$ of the excited state, and both
fields $W$ and $W^{\prime}$ are exactly on resonance. In this case,
the steady-state coherence $\rho_{1a,1b}^e$ between levels $| 1a
\rangle$ and $| 1b \rangle$ is given by
\begin{equation}
\rho_{1a,1b}^e = -\frac{\Omega_W^* \Omega_{W^{\prime}}}{|\Omega_W|^2
+ |\Omega_{W^{\prime}}|^2} \;, \label{rhoe}
\end{equation}
with
\begin{subequations}
\begin{align}
\Omega_W &= \frac{id_{2,1a} {\cal E}_W e^{ik_W z}}{\hbar}\;, \\
\Omega_{W^{\prime}} &= \frac{id_{2,1b}{\cal E}_{W^{\prime}}
e^{i\vec{k}_{W^{\prime}}\cdot \vec{r}}}{\hbar}\;,
\end{align}
\label{OmegasW}
\end{subequations}
the Rabi frequencies related to fields $W$ and $W^{\prime}$,
respectively, and $d_{i,j}$ the dipole moment between levels $i$ and
$j$. In the same situation, the populations of the ground states are
given by
\begin{subequations}
\begin{align}
\rho_{1a,1a}^e &= \frac{|\Omega_W|^2}{|\Omega_W|^2 + |\Omega_{W^{\prime}}|^2} \;, \\
\rho_{1b,1b}^e &= \frac{|\Omega_{W^{\prime}}|^2}{|\Omega_W|^2 +
|\Omega_{W^{\prime}}|^2} \;.
\end{align}
\label{rhoii}
\end{subequations}

If we wait a time $t_s >> 1/\Gamma_{22}$ after the fields $W$ and
$W^{\prime}$ are turned off, the only non-zero elements of the
density matrix $\hat{\rho}^s$ specifying the state of the system at
$t_s$ are
\begin{subequations}
\begin{align}
\rho_{1a,1a}^s &= \frac{1}{2} + \frac{1}{2}(\rho_{1a,1a}^e - \rho_{1b,1b}^e) \;, \\
\rho_{1b,1b}^s &= \frac{1}{2} - \frac{1}{2}(\rho_{1a,1a}^e - \rho_{1b,1b}^e) \;, \\
\rho_{1a,1b}^s &= \rho_{1a,1b}^e e^{-\gamma t_s} \;.
\end{align}
\label{rhos}
\end{subequations}
This gives then the state of the atomic ensemble, with the stored
coherence grating, right before the readout fields are turned on at
an arbitrary time $t_s >> 1/\Gamma_{22}$.

\subsection{Pulse pair generation}
\label{generation}

The reading process starts once fields $R$ and $R^{\prime}$ are
turned on. As discussed above, these two fields propagate in
opposite directions along the $z$ axis and their electric fields are
given by
\begin{subequations}
\begin{align}
\vec{E}_R &= {\cal E}_R e^{i(-k_R z - \omega_R t)} \hat{\sigma}^{-} \;, \\
\vec{E}_{R^{\prime}} &= {\cal E}_{R^{\prime}} e^{i(k_{R^{\prime}}z -
\omega_{R^{\prime}} t)} \hat{\sigma}^{+} \;,
\end{align}
\label{camposR}
\end{subequations}
with ${\cal E}_R$ and ${\cal E}_{R^{\prime}}$ their constant
amplitudes, $\omega_R$ and $\omega_{R^{\prime}}$ the laser
frequencies, and $\vec{k}_R$ and $\vec{k}_{R^{\prime}}$ specifying their
wave-vectors along the $z$ axis. We have then $\vec{k}_{R^{\prime}}
= \vec{k}_W = -\vec{k}_R$. Defining the Rabi frequencies $\Omega_R$
and $\Omega_{R^{\prime}}$ analogously as in Eqs.~(\ref{OmegasW}a)
and~(\ref{OmegasW}b), we can then write the following Bloch
equations for the time evolution of the system in the readout
process
\begin{subequations}
\begin{align}
\frac{d\rho_{1a,1a}}{dt} &= \left[ -\Omega_{R^{\prime}} \,\sigma_{1a,2} + c.c.\right] \nonumber \\ & \hspace{1cm} + \frac{\Gamma_{22}}{2}(1-\rho_{1a,1a}-\rho_{1b,1b}) \;, \\
\frac{d\rho_{1b,1b}}{dt} &= \left[ -\Omega_R \,\sigma_{1b,2} + c.c.\right] \nonumber \\ & \hspace{1cm}  + \frac{\Gamma_{22}}{2}(1-\rho_{1a,1a}-\rho_{1b,1b})\;,\\
\frac{d\sigma_{1a,2}}{dt} &= -\Omega_{R^{\prime}}^* (1 - 2\rho_{1a,1a}-\rho_{1b,1b}) + \Omega_R^*\rho_{1a,1b} \nonumber \\ & \hspace{1cm}  - \frac{\Gamma_{22}}{2}\sigma_{1a,2} \;,\\
\frac{d\sigma_{1b,2}}{dt} &= -\Omega_R^* (1 - 2\rho_{1b,1b}-\rho_{1a,1a}) + \Omega_{R^{\prime}}^* \rho_{1b,1a} \nonumber \\ & \hspace{1cm}  - \frac{\Gamma_{22}}{2}\sigma_{1b,2} \;,\\
\frac{d\rho_{1a,1b}}{dt} &= -\Omega_{R^{\prime}}^* \,\sigma_{2,1b} -
\Omega_R \,\sigma_{1a,2} \;,
\end{align}
\label{Bloch_R1}
\end{subequations}
where we neglected the decoherence rate $\gamma$ of the ground
state, considering a fast readout with $\Omega_R,
\Omega_{R^{\prime}} >> \gamma$. The laser fields are again at exact
resonance, and $\sigma_{1a,2} = \rho_{1a,2}
e^{-\omega_{R^{\prime}}t}$ and $\sigma_{1b,2} = \rho_{1b,2}
e^{-\omega_Rt}$ are the slowly varying coherences.

In order to obtain the temporal shape of the retrieved pulses, we
need to analytically solve this system of eight equations for both
$\sigma_{1a,2}(t)$ and $\sigma_{1b,2}(t)$, considering the initial
conditions as the stored state discussed previously, i.e.,
Eqs.~(\ref{rhos}a) to~(\ref{rhos}c) plus $\sigma_{1a,2}(0) =
\sigma_{1b,2}(0) = 0$. An important step for such analytical
solution is the definition of the following new variables:
\begin{subequations}
\begin{align}
Q &= \Omega_{R}\sigma_{1b,e}\;, \\
P &= \Omega_{R^{\prime}} \sigma_{1a,2}\;, \\
T &= \Omega_{R^{\prime}}\Omega_R^*\rho_{1a,1b} \;,
\end{align}
\label{QPT}
\end{subequations}
and
\begin{subequations}
\begin{align}
Q_r &= Q + Q^* \;, \\
Q_i &= Q - Q^* \;, \\
P_r &= P + P^* \;, \\
P_i &= P - P^* \;, \\
T_r &= T + T^* \;, \\
T_i &= T - T^* \;.
\end{align}
\label{QrtoTi}
\end{subequations}
In terms of this last set of variables, Eqs.~(\ref{Bloch_R1}a)
to~(\ref{Bloch_R1}e) split in two decoupled systems, which we call
just ``System 1'':
\begin{subequations}
\begin{align}
\frac{dP_i}{dt} &= T_i - \frac{\Gamma_{22}}{2}P_i \;,\\
\frac{dQ_i}{dt} &= -T_i - \frac{\Gamma_{22}}{2} Q_i \;,\\
\frac{dT_i}{dt} &= |\Omega_{R^{\prime}}|^2 Q_i - |\Omega_R|^2 P_i
\;;
\end{align}
\label{Bloch_R2a}
\end{subequations}
and ``System 2'':
\begin{subequations}
\begin{align}
\frac{d\rho_{1a,1a}}{dt} &= - P_r + \frac{\Gamma_{22}}{2}(1-\rho_{1a,1a}-\rho_{1b,1b}) \;, \\
\frac{d\rho_{1b,1b}}{dt} &= -Q_r + \frac{\Gamma_{22}}{2}(1-\rho_{1a,1a}-\rho_{1b,1b})\;,\\
\frac{dP_r}{dt} &= -2 |\Omega_{R^{\prime}}|^2 (1 - 2\rho_{1a,1a}-\rho_{1b,1b}) + T_r - \frac{\Gamma_{22}}{2}P_r \;,\\
\frac{dQ_r}{dt} &= -2|\Omega_R|^2 (1 - 2\rho_{1b,1b}-\rho_{1a,1a}) + T_r - \frac{\Gamma_{22}}{2} Q_r \;,\\
\frac{dT_r}{dt} &= -|\Omega_{R^{\prime}}|^2 Q_r - |\Omega_R|^2 P_r
\;.
\end{align}
\label{Bloch_R2b}
\end{subequations}

The solution of System 1 is quite straightforward. Considering the
initial conditions and defining
\begin{equation}
I_t = I_R + I_{R^{\prime}} = \frac{8(|\Omega_R|^2 +
|\Omega_{R^{\prime}}|^2)}{\Gamma_{22}^2}\;, \label{It}
\end{equation}
an adimensional quantity roughly proportional to the total intensity
of light exciting the atoms, we obtain
\begin{subequations}
\begin{align}
P_i(t) &= T_i(0)f_r(t)/\,\Gamma_{22} \;,\\
Q_i(t) &= -T_i(0)f_r(t)/\,\Gamma_{22} \;,
\end{align}
\label{Bloch_R3a}
\end{subequations}
with
\begin{equation}
f_r(t) = \frac{e^{-\Gamma_{22}t/4}\,\mbox{sinh}\left(
\sqrt{1-2I_t}\,\,\Gamma_{22} t/4 \right) }{\sqrt{1-2I_t}\,\,/4} \;.
\label{Bloch_20d}
\end{equation}
Note that $I_R$ and $I_{R^{\prime}}$ give the respective intensity
of the fields in units of the saturation intensity of the
transitions as defined in Ref.~\cite{Steck}. In this way,
$I_t$ is the total intensity of the field in units of the respective
saturation intensities.

The solution of System 2 is more elaborated~\cite{system2}, and we
have to define yet another adimensional quantity
\begin{equation}
I_d = I_R - I_{R^{\prime}}\;,
\end{equation}
proportional now to the difference between the intensities of fields
$R$ and $R^{\prime}$. Taking into account the initial conditions, we have
then
\begin{subequations}
\begin{align}
P_r(t) &= \Gamma_{22} f_r(t) \left[ \frac{I_R I_R^{\prime}}{4 \, I_t}(\rho_{1a,1a}^e - \rho_{1b,1b}^e)+\frac{I_d}{I_t}\frac{T_r(0)}{\Gamma_{22}^2}\right] \nonumber \\
           & \;\; + \Gamma_{22} g_r(t)\frac{I_{R^{\prime}}}{8 \, I_t} \left[ I_t - I_d(\rho_{1a,1a}^e - \rho_{1b,1b}^e)+16\frac{T_r(0)}{\Gamma_{22}^2} \right]\;,\\
Q_r(t) &= -\Gamma_{22} f_r(t) \left[ \frac{I_R I_R^{\prime}}{4 \, I_t}(\rho_{1a,1a}^e - \rho_{1b,1b}^e)+\frac{I_d}{I_t}\frac{T_r(0)}{\Gamma_{22}^2}\right] \nonumber \\
           & \;\; + \Gamma_{22} g_r(t)\frac{I_{R}}{8 \, I_t} \left[ I_t - I_d(\rho_{1a,1a}^e - \rho_{1b,1b}^e)+16\frac{T_r(0)}{\Gamma_{22}^2} \right]\;,
\end{align}
\label{Bloch_R3b}
\end{subequations}
with
\begin{align}
g_r(t) =& \frac{2r_3+2r_1+ 1}{2(r_2-r_1)(r_2-r_3)} \left( e^{r_1\Gamma_{22}t} - e^{r_2\Gamma_{22}t}\right) \nonumber \\
        & \;\; + \frac{2r_2+2r_1+ 1}{2(r_3-r_1)(r_3-r_2)} \left( e^{r_1\Gamma_{22}t} - e^{r_3\Gamma_{22}t}\right) \;.
\end{align}
The coeficients $r_1$, $r_2$, and $r_3$ are given by
\[
r_1 = s + v -\frac{1}{2} \;,
\]
\[
r_2 = -\frac{(s+v)}{2} - \frac{1}{2} + \frac{\sqrt{3}}{2} (s-v) i \;,
\]
\[
r_3 = -\frac{(s+v)}{2} - \frac{1}{2} - \frac{\sqrt{3}}{2} (s-v) i \;,
\]
with
\[
s = \sqrt[3]{\frac{I_t}{16}+\sqrt{\frac{1}{8}\left(\frac{I_t}{3}-\frac{1}{6}\right)^3 + \frac{I_t^2}{16^2}}} \;,
\]
\[
v = \sqrt[3]{\frac{I_t}{16}-\sqrt{\frac{1}{8}\left(\frac{I_t}{3}-\frac{1}{6}\right)^3 + \frac{I_t^2}{16^2}}} \;.
\]
In the same way as $f_r(t)$, $g_r(t)$ depends then only on $I_t$ and $\Gamma_{22}$.

With $P_i$, $P_r$, $Q_i$, and $Q_r$, it is then possible to invert
Eqs.~\eqref{QPT} and~\eqref{QrtoTi} and obtain directly
\begin{subequations}
\begin{align}
\sigma_{1a,2}(t) =& \frac{P_r(t) + P_i(t)}{2 \Omega_{R^{\prime}}}\;, \\
\sigma_{1b,2}(t) =& \frac{Q_r(t) + Q_i(t)}{2 \Omega_R}\;,
\end{align}
\end{subequations}
which gives then the medium polarization responsible for the
emission of the $D$, $D^{\prime}$ pulse pair.

\subsubsection{Equal populations in the ground states}
\label{equalros}

An important situation to be analyzed is when $|\Omega_W| =
|\Omega_{W^{\prime}}|$, which leads to $\rho_{1a,1a}^e =
\rho_{1b,1b}^e$. This case corresponds to a maximum visibility of
the stored coherence grating, and to simpler expressions for both
$\sigma_{1a,2}$ and $\sigma_{1b,2}$:
\begin{subequations}
\begin{align}
\sigma_{1a,2}(t) =& \; -i g_r(t)\frac{|\Omega_{R^{\prime}}|}{2\Gamma_{22}}e^{-i\vec{k}_{R^{\prime}}\cdot \vec{r}} \nonumber \\
             & \; -i \frac{|\Omega_R|e^{-\gamma t_s}}{2 I_t \Gamma_{22}} \left[ f_r(t)I_R + g_r(t) I_{R^{\prime}} \right] \nonumber \\
                  & \;\;\;\; \times e^{-i(\vec{k}_R+\vec{k}_W-\vec{k}_{W^{\prime}})\cdot \vec{r}} \nonumber \\
             & \; +i \frac{I_{R^{\prime}} |\Omega_R|e^{-\gamma t_s}}{2 I_t \Gamma_{22}} \left[ f_r(t) - g_r(t) \right] \nonumber \\
                  & \;\;\;\; \times e^{-i(2\vec{k}_{R^{\prime}}-\vec{k}_R-\vec{k}_W+\vec{k}_{W^{\prime}})\cdot \vec{r}} \;, \\
\sigma_{1b,2}(t) =& \; -i g_r(t)\frac{|\Omega_{R}|}{2\Gamma_{22}}e^{-i\vec{k}_{R}\cdot \vec{r}} \nonumber \\
             & \; -i \frac{|\Omega_{R^{\prime}}|e^{-\gamma t_s}}{2 I_t \Gamma_{22}} \left[ f_r(t)I_{R^{\prime}} + g_r(t) I_{R}\right] \nonumber \\
                  & \;\;\;\; \times e^{-i(\vec{k}_{R^{\prime}}-\vec{k}_W+\vec{k}_{W^{\prime}})\cdot \vec{r}} \nonumber \\
             & \; +i \frac{I_R |\Omega_{R^{\prime}}| e^{-\gamma t_s}}{2 I_t \Gamma_{22}} \left[ f_r(t) - g_r(t) \right] \nonumber \\
                  & \;\;\;\; \times e^{-i(2\vec{k}_R-\vec{k}_{R^{\prime}}+\vec{k}_W-\vec{k}_{W^{\prime}})\cdot \vec{r}}\;,
\end{align}
\label{sigmas}
\end{subequations}
where we used $\Omega_X = i |\Omega_X|e^{i\vec{k}_X\cdot\vec{r}}$,
with $X$ being $R$, $R^{\prime}$, $W$, or $W^{\prime}$. We also
substituted $\rho_{1a,1b}(0)=\rho_{1a,1b}^s=e^{-\gamma t_s}/2$, as
is the case once $|\Omega_W| = |\Omega_{W^{\prime}}|$.

As discussed in more detail in Ref.~\cite{Moretti2008}, the
second term in Eq.~(\ref{sigmas}a) is the one responsible for the
$D$ field back propagating with respect to $W^{\prime}$.
Correspondingly, the $D^{\prime}$ field comes from the second term
in Eq.~(\ref{sigmas}b), which generates a field propagating in the
direction of $W^{\prime}$. The first term on these two equations
correspond to stimulated emission on the respective transitions,
contributing to fields on the directions of $R$ and $R^{\prime}$.
The third term, for the geometrical configuration of the beams we
consider here, is not phase-matched and does not result in a
propagating field.

For the sake of the following discussions, we only need to consider
then the electric field amplitudes of $D$ and $D^{\prime}$:
\begin{subequations}
\begin{align}
E_D(t) \propto& \; \frac{|\Omega_R|}{I_t\Gamma_{22}} \left[ f_r(t)I_R + g_r(t)I_{R^{\prime}}\right]  \;, \\
E_{D^{\prime}}(t) \propto& \;
\frac{|\Omega_{R^{\prime}}|}{I_t\Gamma_{22}} \left[ f_r(t)
I_{R^{\prime}} + g_r(t) I_{R} \right] \;.
\end{align}
\label{EDs}
\end{subequations}
For detectors with a fast response, the respective measured
signals~\cite{Moretti2008} should be given by
\begin{subequations}
\begin{align}
S_D(t) =& \; A_D \frac{I_R}{I_t^2} \left| f_r(t)I_R + g_r(t) I_{R^{\prime}}\right|^2  \;, \\
S_{D^{\prime}}(t) =& \; A_{D^{\prime}}\frac{I_{R^{\prime}}}{I_t^2}
\left| f_r(t) I_{R^{\prime}} + g_r(t) I_{R} \right|^2 \;,
\end{align}
\label{Ss}
\end{subequations}
with $A_D$ and $A_{D^{\prime}}$ two proportionality constants.
Equations~(\ref{Ss}a) and~(\ref{Ss}b) will be directly compared to
the experimental results of Sec.~\ref{experiment}, and theirs
applicability will be discussed there in more detail. Examples of theoretical pulse shapes can be found in Fig.~\ref{fig5}. A physical
situation where only the $f_r(t)$ function determines the temporal
dynamics was explored in Ref.~\cite{Moretti2008}.

\subsubsection{Energy in the diffracted modes}

Another important quantity to be investigated is the retrieved
energy in modes $D$ and $D^{\prime}$, since these are directly
related to the retrieval efficiency in the reading process for the
information stored in the atomic ensemble~\cite{Laurat2006}. They
will be given by~\cite{Moretti2008}
\begin{subequations}
\begin{align}
U_D =& \int_0^{\infty}S_D(t)dt \nonumber \\ =& \; B_D \frac{I_R}{I_t^2} \int_0^{\infty} \left| f_r(t)I_R + g_r(t) I_{R^{\prime}}\right|^2 dt \;, \\
U_{D^{\prime}} =& \int_0^{\infty}S_{D^{\prime}}(t)dt \nonumber \\ =&
\; B_{D^{\prime}}\frac{I_{R^{\prime}}}{I_t^2} \int_0^{\infty}\left|
f_r(t) I_{R^{\prime}} + g_r(t) I_{R} \right|^2dt \;,
\end{align}
\label{Us}
\end{subequations}
with $B_D$ and $B_{D^{\prime}}$ two new proportionality constants.
In this way, the total energy retrieved will be
\begin{equation}
U_T(t) = U_D + U_{D^{\prime}} \;. \label{UT}
\end{equation}
All these quantities ($U_D$, $U_{D^{\prime}}$, and $U_T$) will be
experimentally investigated in the next section. For the following calculations and analysis, we are going to consider completely symmetrical conditions for the generation of $D$ and $D^{\prime}$, so that $B_D = B_{D^{\prime}}$.

\begin{figure}[ht]
\hspace{0.0cm}\includegraphics[width = 8.5 cm,angle=0]{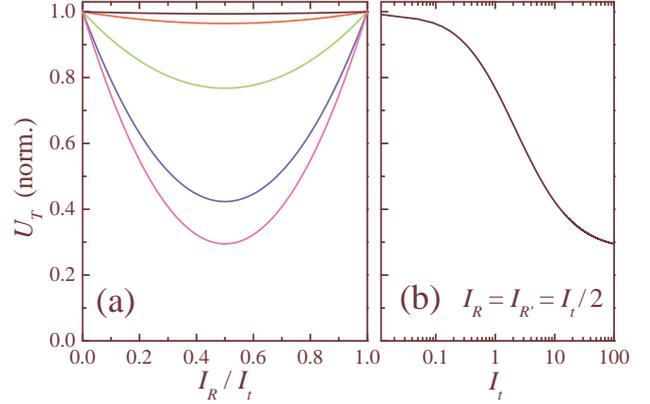}
\vspace{-2.5cm}
\caption{(a) Normalized total energy in the diffracted modes $D$,$D^{\prime}$ as a function of the $R$ beam intensity $I_R$, in a situation where the $R^{\prime}$ beam intensity is $I_{R^{\prime}} = I_t - I_R $. $I_t$ is then kept constant for each curve, with its value taken, from top to botton, as 0.01, 0.1, 1.0, 10, and 100, respectively. Note that the minimum always occurs at $I_R = I_{R^{\prime}} = I_t /2$. The curves are normalized by their value at $I_R = 0$. (b) Total energy in the diffracted modes at the minimum as a function of the total intensity $I_t$. For each $I_t$, the curve is normalized by the $U_T$ value with $I_R = 0$ and $I_{R^{\prime}}=I_t$.} 
\label{fignova}
\end{figure}

In general, $U_T$ depends only on the three intensities $I_t$, $I_R$, and $I_{R^{\prime}}$. The way we chose to explore this quantity was to fix the total intensity and change only the relative ratio between the two reading beams, i.e., we studied $U_T$ as a function of $I_R$ in a situation where $I_{R^{\prime}} = I_t - I_R$. We obtain then graphs as shown in Fig.~\ref{fignova}a, where we plotted $U_T$ as a function of $I_R$ for five different values of $I_t$, going from 0.01 to 100. These plots clearly demonstrate that the total diffracted energy presents a minimum at $I_R = I_{R^{\prime}} = I_t/2$ once $I_t$ gets large enough ($I_t > 1$). This behavior indicates then a specific kind of saturation in this system, which prevents the whole retrieved energy to be evenly distributed among these two modes for large total intensities.

\vspace*{1.8cm}

From the theoretical expressions deduced above, this behavior comes as a consequence of the fact that $f_r(t) \approx g_r(t)$ in the limit $I_t << 1$. If we make $f_r = g_r$ in Eqs.~\eqref{Us}, it is straightforward to see that $U_T$ becomes independent of the relative ratio between $I_R$ and $I_{R^{\prime}}$ and, consequently, does not present a minimum anymore. On the other hand, in the limit $I_t > 1$ the function $|g_r(t)|^2$ becomes consistently smaller than $|f_r(t)|^2$, resulting in the overall dependence of $U_T$ with $I_R, \,I_{R^{\prime}}$. The minimum formation as $I_t$ is increased can be better visualized in Fig.~\ref{fignova}b, in which the solid line represents the minimum $U_T(I_R = I_{R^{\prime}} = I_t/2)$ normalized by the value of the total diffracted energy when all the reading power is in just one reading beam, i.e., normalized by $U_T(I_R = 0, I_{R^{\prime}}=I_t)$ as in Fig.~\ref{fignova}a. We see then that we go from no minimum at $I_t << 1$ to a value asymptotically approaching 0.277 at $I_t >> 1$. The interpretation of this specific value for the asymptotic limit of the minimum is still under investigation.

At first sight, the dependence of the total diffracted energy with the relative intensities of the reading beams may be attributed to any process that can affect the diffraction efficiency in a similar way, and which is maximized when the two reading intensities are equal. One such process could be spontaneous emission, which might erase the stored information, therefore reducing the conversion efficiency of energy of the beams $R$,$R^{^{\prime}}$ into the beams $D$,$D^{\prime}$. However, although the excited state population depends locally on the relative reading beam intensities, one can easily show (Appendix~\ref{ApA}) that the total amount of spontaneous emission in the atomic ensemble depends only on the total intensity $I_t$ and cannot be responsible for the observed dependence. Similarly, a calculation of the total energy emitted by the atomic ensemble in the form of stimulated emission also leads to an expression which depends only on $I_t$ (see also Appendix~\ref{ApA}), not presenting a minimum at $I_R = I_{R^{\prime}} = I_t/2$. Therefore, we are left with the coherent process associated with the third terms of Eqs.~\eqref{sigmas}. These terms represent a spatially dependent polarization induced simultaneously by both reading beams and are responsible by a portion of the local atomic excitation. Although they represent a process which is not phase matched for the reading beam configuration of Fig.~\ref{fig1} and, therefore, do not lead to the generation of a coherent beam, their excitation leads to a decrease in the conversion efficiency of the reading beams into the diffracted $D$,$D^{\prime}$ modes. Furthermore, it can also be shown (Appendix~\ref{ApB}) that the total irradiated energy associated with these terms scales with the product of the two reading beam intensities, reaching a maximum when $I_R = I_{R^{\prime}} = I_t/2$. Thus, we attribute the predicted minimum as being due to excitation of these non-phase-matched optical coherences in the medium.  

\vspace{-0.5cm}
\section{Experiment}
\label{experiment} 

For the experiment we employ a sample of cold
cesium atoms obtained from a magneto-optical trap (MOT), with the trapping beams detuned
about $12\,$MHz from the cycling transition
$6S_{1/2}(F=4)\rightarrow 6P_{3/2}(F^{\prime }=5)$, and the
recycling repumping beam resonant with the open transition
$6S_{1/2}(F=3)\rightarrow 6P_{3/2}(F^{\prime }=3)$. Initially
the atoms are prepared in the lower hyperfine ground state by
switching off the repumping beam for about $1\,$ms and waiting for an
optical pumping period (induced by the non-resonant trapping beams)
of about $50\,\mu$s after which most of the atoms are pumped into the
$6S_{1/2}(F=3)$ ground state. During this optical pumping period the
MOT quadrupole magnetic field is also switched off. We use three
pairs of orthogonal Helmholtz coils to compensate for residual
magnetic fields. The light storage experiment is performed using
light from an external cavity diode laser (ECDL) locked on the
cesium closed transition $6S_{1/2}(F=3)\rightarrow
6P_{3/2}(F^{\prime }=2)$. 

\begin{figure}[ht]
\hspace{0.2cm}\includegraphics[width =8.3 cm,angle=0]{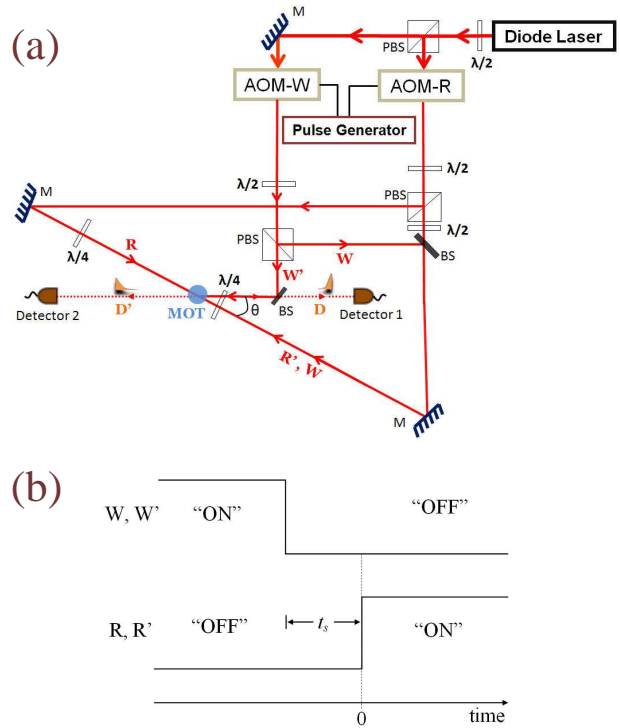}
\vspace{-1.7cm}
\caption{(a) Setup for the simultaneous detection of the $D$ and $D^{\prime}$ pulses. Main components: external cavity diode laser; M, mirror; BS, beam splitter; PBS, polarizing beam splitter; $\lambda/2$, half-wave plate; $\lambda/4$, quarter-wave plate; AOM, acoustic-optical modulator. After detection, both signals are sent to a four-channel 400~MHz oscilloscope. (b) Time sequence for writing ($W$,$W^{\prime}$) and reading ($R$,$R^{\prime}$) fields. $t_s$ is the storage time counted from the turn off of $W$,$W^{\prime}$ up to the turn on of $R$,$R^{\prime}$. The origin of the timescale is the one used in Figs.~\ref{fig3} and~\ref{fig5}.}
\label{fig2}
\end{figure}

The main parts of the experimental setup are describe in Fig.~\ref{fig2}a. In order to control the intensities of the
incident writing beams $W$ and $W^{\prime}$ we use a pair of
acoustic-optical modulators (AOM-W) operating in opposite diffraction
orders so to keep the laser frequency resonant with the atomic
transition. Similarly, another pair of AOMs (AOM-R) allowed for the
independent control of the intensities of the reading beams $R$ and
$R^{\prime}$ in relation to the writing beams. As shown in Fig.~\ref{fig2}a, 
half-waveplates and polarizing beam splitters
allowed us to control of the relative intensities of the writing
beams as well as of the reading beams. The two writing beams $W$ and
$W^{\prime}$ pass through a quarter-waveplate so to acquire opposite
circular polarizations and are incident in the MOT forming a small
angle $\theta = 10^{-3}\,$rad. 

\vspace*{-0.4cm}

The writing pulses are applied
for a sufficiently long period, approximately $40\,\mu$s, so to
allow for the creation of the stationary ground-state Zeeman coherence
grating. The two counter-propagating reading beams $R$ and
$R^{\prime}$ are switched on after a controllable storage time $t_s$
(fixed in $2\,\mu$s in our case) measured from the instant of the
turning off of the writing beams, Fig.~\ref{fig2}b. As can be seen in the experimental
scheme depicted in Fig.~\ref{fig2}a, the reading beams also have opposite
circular polarizations, with the $R^{\prime}$ beam having the same polarization and propagating in the same
direction as the writing $W$ beam. Under these
conditions, we were able to retrieve the two counterpropagating
Bragg diffracted pulses, labeled $D$ and $D^{\prime}$, which are
detected by two photodiodes. It is worth mentioning that we
have verified that the two retrieved pulses also have opposite
circular polarizations, consistent with the conservation of angular
momentum in the whole process.

\begin{figure}[ht]
\hspace{0.0cm}\includegraphics[width =6.5 cm,angle=0]{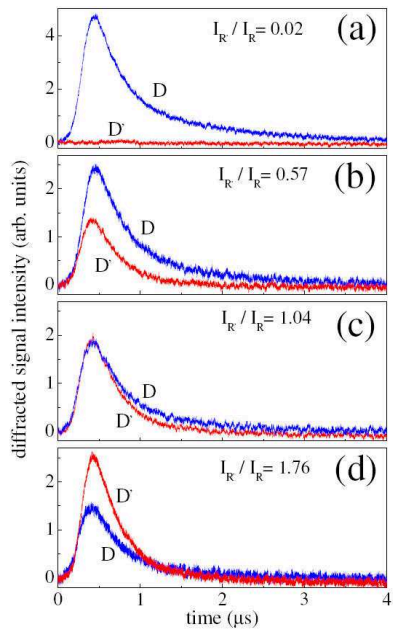}
\vspace{-0.1cm}
\caption{Diffracted signals as a function of time for different relations between the intensities of the reading fields: (a) $I_{R^{\prime}}/I_R = 0.02$, (b) $I_{R^{\prime}}/I_R = 0.57$, (c) $I_{R^{\prime}}/I_R = 1.04$, and (d) $I_{R^{\prime}}/I_R = 1.76$. The intensity of the writing fields, $W$ and $W^{\prime}$, is $1.8\,\mbox{mW/cm}^2$ each. The total intensity of the reading fields is the same, $4.5\,\mbox{mW/cm}^2$, for each frame.} 
\label{fig3}
\end{figure}

In Fig.~\ref{fig3} we show the temporal shape of the retrieved pulses for different
intensity ratios of the reading beams $R$ and $R^{\prime}$. These results demonstrate that we can control, in a straightforward manner, the relative intensity of the two pulses. However, they cannot be directly compared to the theory of the previous section, since we employed two relatively slow detectors (PDA36A from Thorlabs) in our experiments. Their response time was about 0.5~$\mu$s, which gives then a lower bound for the measured temporal width of the pulses. 

\vspace*{-0.2cm}

\begin{figure}[ht]
\hspace{0.0cm}\includegraphics[width =8.5 cm,angle=0]{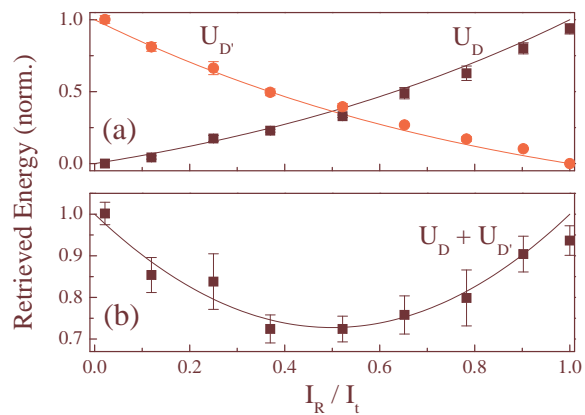}
\vspace{-2.6cm}
\caption{Energy of the diffracted pulses, $D$ and $D^{\prime}$, as a function of the normalized intensity of field $R$. The squares (circles) correspond to the experimental points for $D$ ($D^{\prime}$). In the experiment, the total reading intensity is kept constant and equal to $4.5\,\mbox{mW/cm}^2$. The intensity is $1.8\,\mbox{mW/cm}^2$ for both writing fields. The solid curves are the corresponding theoretical results from Eq.~\eqref{Us}, with $I_t = 1.3$. The error bars were obtained from three independent measurements under the same conditions.} 
\label{fig4}
\end{figure}

With this in mind, the first experimental results we compared to the theory were the ones for the energy of the pulses as a function of the relative reading intensity, Fig.~\ref{fig4}. This energy depends on the integral of the temporal shape of the pulses, which we expect to be less sensitive to the detector's time response. Figure~\ref{fig4}a plots then the energy extracted in each of the diffracted fields, $U_D$ and $U_{D^{\prime}}$, as a function of the intensity of field $R$ normalized by the total intensity $I_t$. The normalization of the $U_T$ values is done by dividing them by the value of $U_T$ with $I_R = 0$ and $I_{R^{\prime}} = I_t$. For the normalization of the experimental values in the plots of Fig.~\ref{fig4}, we have carefully accounted for the 50\% intensity loss in the retrieved $D$ beam caused by the extra beam splitter in its pathway as well as for all other losses along the paths of both $D$,$D^{\prime}$ beams. The two writing fields have the same intensity ($1.8\,\mbox{mW/cm}^2$), so we can compare the experimental results to the simplified theoretical expressions of Sec.~\ref{equalros}. The theoretical curves in Fig.~\ref{fig4} were generated by adjusting $I_t$ in order to obtain the best fit to the experimental points. We considered then $I_t = 1.3$ in units of the saturation intensity of the transitions [see Eq.~\eqref{It} and Ref.~\cite{Steck}], considered here to have the same value for both transitions. In the experiment, we employed a total intensity of $4.5\,\mbox{mW/cm}^2$. The theoretical value is consistent with the experimental one, as being just above the saturation intensity. Notice, however, that a more direct comparison of the two values is not possible under our simplified three-level model, since we do not take into account the whole Zeeman structure of the real atom. 

\vspace*{1.8cm}

The measurements in Fig.~\ref{fig4}a demonstrate that the total energy extracted by the reading process presents a minimum when the two reading fields have the same intensity, as expected from the theory introduced above. This is presented most clearly in Fig.~\ref{fig4}b, a plot of the sum of the two curves in Fig.~\ref{fig4}a. We observe then a reduction of about 30\% in $U_D + U_{D^{\prime}}$ when $I_R = I_{R^{\prime}}$. The relatively simple theory employed here reproduces quite well this behavior, indicating that the relevant physical process behind it is properly taking into account by the theory.

\begin{figure}[ht]
\hspace{0.0cm}\includegraphics[width =7.7 cm,angle=0]{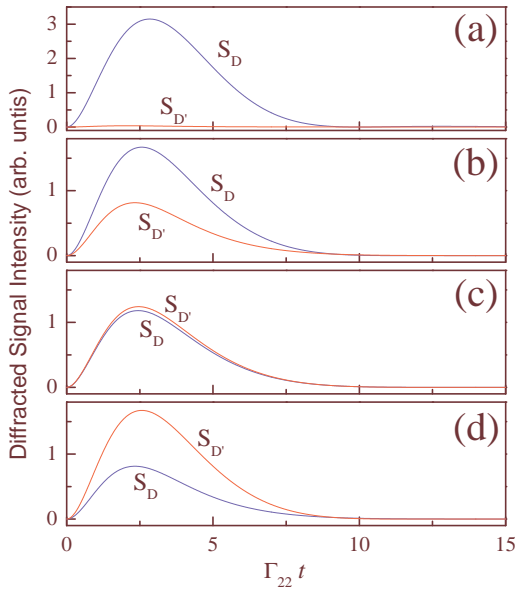}
\vspace{-2.1cm}
\caption{Diffracted signals as a function of time according to Eq.~\eqref{Ss} and for different ratios of the reading intensity: (a) $I_{R^{\prime}}/I_R = 0.02$, (b) $I_{R^{\prime}}/I_R = 0.57$, (c) $I_{R^{\prime}}/I_R = 1.04$, and (d) $I_{R^{\prime}}/I_R = 1.76$. The total intensity, normalized by $\Gamma_{22}$, is kept constant and equal to $I_t = 1.3$.} 
\label{fig5}
\end{figure}

As discussed in connection to Fig.~\ref{fig3}, the measurements of the experimental pulse shapes are limited to a minimum temporal width of about $0.5\,\mu$s due to our detector's response time. In this way, the theoretical pulse shapes for fast detectors, as given by $S_D$ and $S_{D^{\prime}}$ in Eq.~\eqref{Ss}, cannot be directly compared to the experimental curves. Even though, the theory properly describes the split of the extracted diffracted fields among the two modes, $D$ and $D^{\prime}$. In order to illustrate the kind of pulse shapes obtained from such theory under the conditions of Fig.~\ref{fig4}, we provide in Fig.~\ref{fig5} the pulse shapes of $D$ and $D^{\prime}$ for each of the intensity relations of Figs.~\ref{fig3}a to~\ref{fig3}d.

\vspace*{0.0cm}

\vspace{-0.0cm}
\section{Conclusions}
\label{conclusions}

In the present work we reported a detailed study of the process of pulse pair generation from a coherence grating stored in an atomic ensemble. We showed that it is possible to generate pulse pairs with arbitrary relative amplitudes depending on the intensity relation of the two beams employed in the reading process, where the stored grating is mapped into the two diffracted fields. We also observed the formation of a minimum for the total extracted energy when the two reading beams have the same intensity. We provided a simple, analytical theory that accounted for all these features observed in the experiment. From this theory, we obtain that such minimum formation is the result of saturation of the system by the reading fields, which excite other polarization modes that do not lead to diffracted, propagating fields. 

From a more general perspective, the process discussed above provides a continuous interpolation between two more well known reading processes, the ones based on Delayed-Four-Wave-Mixing~\cite{Moretti2008,Matsukevich2005} and EIT~\cite{Liu01,Phillips01}, respectively. This general pulse pair generation process may be used as a means, for example, to deterministically generate complex, correlated pairs of optical fields. A study of the correlation properties of these pulse pairs is currently in progress. It may also be employed to distribute the information present originally in a single optical field among two other fields, after its storage and possible manipulation in the atomic ensemble. This may be an important tool to explore new possibilities in quantum networks.

\bigskip
\noindent
We are grateful to Yareni Ayala for experimental assistence in an early stage of this work. This work was supported by CNPq, CAPES, and FACEPE (Brazilian Agencies), being part of the programs PRONEX and INCT-IQ (Instituto Nacional de Ci\^encia e Tecnologia de Informa\c{c}\~ao Qu\^antica).

\appendix

\section{Spontaneous and Stimulated Emissions}
\label{ApA}

The total amount of spontaneous emission at a given time is given by $\Gamma_{22}\rho_{22}^T$, with $\rho_{22}^T(t)$ the total amount of population in the excited state. In order to obtain an expression for it, our first step is the derivation of $\rho_{22} = 1 - \rho_{1a,1a} - \rho_{1b,1b}$ for a particular atom. From Eqs.~(\ref{Bloch_R2b}a) and~(\ref{Bloch_R2b}b), we have
\begin{equation}
\frac{\rho_{22}}{dt} = P_r + Q_r - \Gamma_{22} \rho_{22} \;.
\end{equation}
Using Eqs.~\eqref{Bloch_R3b} under the condition $|\Omega_W| = |\Omega_{W^{\prime}}|$, we then obtain
\begin{equation}
\rho_{22}(t) = \left[ I_t + \frac{16\,T_r(0)}{\Gamma_{22}^2} \right] \frac{\Gamma_{22}e^{-\Gamma_{22}t}}{8}\int_0^t g_r(t^{\prime})e^{\Gamma_{22}t^{\prime}}dt^{\prime} \;.
\end{equation}  
From the definition of $T_r(t)$ with the initial state given by Eq.~(\ref{rhos}c):
\begin{equation}
T_r(0) = - \frac{e^{-\gamma t_s}|\Omega_R||\Omega_{R^{\prime}}|}{2} \cos (k_{R^{\prime}}z +\vec{k}_{W^{\prime}}\cdot\vec{r}) \;.
\end{equation}
When integrating over the whole ensemble [as done to go from Eqs.~\eqref{sigmas} to Eqs.~\eqref{EDs}], such term multiplied by the cossine of a function of the spatial grating averages to zero. The total population in the excited state is then proportional to
\begin{equation}
\rho_{22}^T(t) \propto \frac{\Gamma_{22} I_t e^{-\Gamma_{22}t}}{8}\int_0^t g_r(t^{\prime})e^{\Gamma_{22}t^{\prime}}dt^{\prime} \;.
\end{equation}
Notice that this expression depends only on the total intensity $I_t$ employed in the reading process and not on the relative intensity of $R$, $R^{\prime}$.

The calculation of the energy emitted by the atomic ensemble in the form of stimulated emission is even more straightforward, since it corresponds to the first term on the right in Eqs.~(\ref{sigmas}a) and~(\ref{sigmas}b). Following similar steps as for the derivation of $U_D$, $U_{D^{\prime}}$, we obtain then expressions for the energy emitted by the atomic system in the reading modes:
\begin{subequations}
\begin{align}
U_R &\propto I_R \int_0^{\infty} |g_r(t)|^2 dt \;,\\
U_{R^{\prime}} &\propto I_{R^{\prime}} \int_0^{\infty} |g_r(t)|^2 dt \;.
\end{align}
\end{subequations} 
These result in an expression of the total energy $U_T^{S}$ emitted by stimulated emission,
\begin{equation}
U_T^S = U_R + U_{R^{\prime}} \propto I_t \int_0^{\infty} |g_r(t)|^2 dt \;,
\end{equation} 
that is also independent of the relative intensity of the reading beams.

\section{Third terms of Eqs.~(\ref{sigmas})}
\label{ApB}

In order to understand how the third terms of Eqs.~\eqref{sigmas} may affect the energy diffracted in the $D$,$D^{\prime}$ modes, we can calculate the total energy scattered incoherently by those terms of the induced polarization, or even consider a different experimental situation for which these terms are phase matched (as in the beams configurations of Refs.~\cite{Cardoso2002} and~\cite{Ducloy1985}). In both cases, the energy scattered by each of these polarization components are given by:
\begin{subequations}
\begin{align}
U_{a} &= B_{a} \frac{I_{R^{\prime}}^2 I_R}{I_t^2} \int_0^{\infty} |f_r(t)-g_r(t)|^2 dt \;,\\
U_{b} &= B_{b} \frac{I_{R^{\prime}} I_R^2}{I_t^2} \int_0^{\infty} |f_r(t)-g_r(t)|^2 dt \;,
\end{align}
\end{subequations}
for the transitions starting at level $1a$ and $1b$, respectively. Under usual conditions, we should have completely symmetrical emissions for both transitions and, consequently, $B_{a} = B_{b}$. The total energy related to these terms would be then
\begin{equation}
U_T^{\prime} = B_{a} \frac{I_RI_{R^{\prime}}}{I_t}\int_0^{\infty} |f_r(t)-g_r(t)|^2 dt\;.
\end{equation}
Now we do have a process for which the maximum emitted energy would occur for the same conditions at which a minimum is observed in the total energy of modes $D$, $D^{\prime}$.


\begin{thebibliography}{16}
\bibitem{system2}
In order to solve System 2 analytically, we define $N=P_r + Q_r$ and
$M = P_r - Q_r$ and obtain first a third-order homogeneous ordinary
differential equation for $N(t)$. After solving this equation, we
obtain a second-order non-homogeneous ordinary diferential equation
for $M(t)$ that can be solved directly once we know $N(t)$. From $N$
and $M$, we can then obtain again $P_r(t)$ and $Q_r(t)$.
\end{thebibliography}
\end{document}